\documentclass[10pt,aps,prl,reprint,showpacs,floatfix,citeautoscript,longbibliography,superscriptaddress]{revtex4-1}

\usepackage{times}
\usepackage{graphicx}
\usepackage{float}
\usepackage{amsmath}
\usepackage{xfrac}
\usepackage{upgreek}
\usepackage{dcolumn}
\usepackage{bm}
\usepackage{caption}
\usepackage{ragged2e}
\usepackage[version=3]{mhchem}

\renewcommand{\raggedright}{\leftskip=0pt \rightskip=0pt plus 0cm}

\begin{document}

\renewcommand{\thefootnote}{\fnsymbol{footnote}}

\title{Quantum-well resonances caused by partial confinement in MgO-based magnetic tunnel junctions}

\author{L. N. Jiang}
\affiliation{Beijing National Laboratory for Condensed Matter Physics, Institute of Physics, Chinese Academy of Sciences, Beijing 100190, China}
\author{B. Y. Chi}
\affiliation{Beijing National Laboratory for Condensed Matter Physics, Institute of Physics, Chinese Academy of Sciences, Beijing 100190, China}
\affiliation{Center of Materials Science and Optoelectronics Engineering, University of Chinese Academy of Sciences, Beijing 100049, China}
\author{W. Z. Chen}
\affiliation{Center for Advanced Quantum Studies and Department of Physics, Beijing Normal University, Beijing 100875, China}
\author{X. F. Han}
\email{Corresponding author: xfhan@iphy.ac.cn}
\affiliation{Beijing National Laboratory for Condensed Matter Physics, Institute of Physics, Chinese Academy of Sciences, Beijing 100190, China}
\affiliation{Center of Materials Science and Optoelectronics Engineering, University of Chinese Academy of Sciences, Beijing 100049, China}
\affiliation{Songshan Lake Materials Laboratory, Dongguan, Guangdong 523808, China}

\date{\today}

\begin{abstract}

  Quantum-well resonance is achieved through partial confinement in magnetic tunnel junctions (MTJs), which provides an additional operable degree of freedom to regulate quantum-well levels. Using Al/Fe/MgO/Fe/Al and Ag/Al/Fe/MgO/Fe/Al/Ag MTJs as examples, via first-principles calculations, we demonstrate that the partial confinement of $\Delta_1$ electron at Al/Fe interface and the full confinement at Fe/MgO interface combine to produce quantum-well resonances in Fe. The quantum-well levels of Fe can be periodically adjusted by two degrees of freedom: Fe and Al thickness. The oscillation period obtained from conductance $G_{\uparrow\uparrow}$ is 2.13 ML Fe (9 ML Al), close to 2.25 ML Fe (8.33 ML Al) calculated by bcc-Fe (fcc-Al) band. The combination of long and short periods enables quantum-well levels to be finely adjusted. An ultrahigh optimistic TMR effect of $3.05\times10$$^5$\% is achieved. Our results provides a new path for designing and applying quantum-well resonances in spintronics devices.

  \end{abstract}

\maketitle

 The study of quantum confinement in magnetic tunnel junctions (MTJs) is crucial for the development of high-performance spintronics devices, such as magnetic random access memories (MRAM), magnetic sensors, and quantum information devices. When electrons are confined in ferromagnetic (FM) layer of MTJs, there will be spin-polarized quantum-well effects. We can make one spin-polarized quantum-well level coincides with the Fermi level, while another spin-polarized level does not. It is a powerful way to enhance the tunneling magnetoresistance (TMR) effect \cite{PhysrevB:5484,PhysRevLett:207210,nanolett:805}, which is an important performance parameter of MTJs \cite{PhysLettA:225,JMMM:L231,PhysRevLett:3273,PhysrevB:054416,nm:862,nm:868,apl:112404,apl:082508}, achieving highly efficient information reading in MRAM and highly sensitive magnetic sensors.

 Generally, the quantum confinement in MTJs is obtained through two interfaces with reflectivity of $\sim$100\% (full confinement), such as MgO/Fe/MgO or Cr/Fe/MgO \cite{apl:4381,PhysRevLett:027208,PhysRevLett:087210,PhysRevLett:187202,PhysRevLett:047207,PhysRevLett:157204,advsci:1901438,jpd:443001}. There is a relatively bandgap between Fe and MgO, and the free electrons or the key $\Delta_1$-symmetry electrons in FM layer are almost 100\% reflected back at two interfaces, resulting in quantum-well states. The quantum-well levels can be regulated by changing the thickness of the FM layer. However, the oscillation period of the key $\Delta_1$ electrons near the Fermi level for commonly used FM metals (Fe \cite{PhysrevB:134404}, Co, CoFe) in MTJs is only about 2$\sim$3 monolayers (ML). With such a short period, changing the thickness of one ML will markedly move quantum-well levels, which is difficult to achieve fine regulation.

 When the interface reflectivity is relatively high but less than 100\% (partial confinement), quantum wells may not be generated in common cognition. However, based on inverse photoemission and photoemission studies on quantum confinement in thin metal films, confining interfaces are analogue to mirrors of a Fabry-Perot interferometer \cite{PhysrevB:1540,Science:1709,PhysRevLett:156801}. Two mirrors or interfaces with reflectivity $\textless$100\% can also produce significant interference, with full (partial) confinement generating quantum-well states (resonances) \cite{PhysrevB:1540,Science:1709,PhysRevLett:156801}. Even when the interface reflectivity is reduced to less than 0.2, the quantum-well peaks broaden but can still be observed in the emission spectra of Ag film on Fe \cite{Science:1709}.

 In this Letter, through first-principles calculations of Al/Fe/MgO/Fe/Al and Ag/Al/Fe/MgO/Fe/Al/Ag MTJs, we achieved quantum-well resonances through partial confinement, and obtained an ultrahigh TMR ratio ($3.05\times10$$^5$\%). Compared to the full confinement, the partial confinement has three major advantages when applied to MTJs.
 One is that the $\Delta_1$ wave function does not decay exponentially in the Al layer as in gap materials in full confinement system, which is beneficial for reducing resistance area. The other is that the quantum-well levels in Fe can be regulated by both Fe and Al thickness. These two degrees of freedom, in conjunction with each other, can finely adjust the quantum-well level to fall at the Fermi level. The last one is that the partial confinement can be extended to a large number of structures. These features allow us to design MTJs more flexibly.

\begin{figure}[thb!]
\centering
\includegraphics[width=8.6cm]{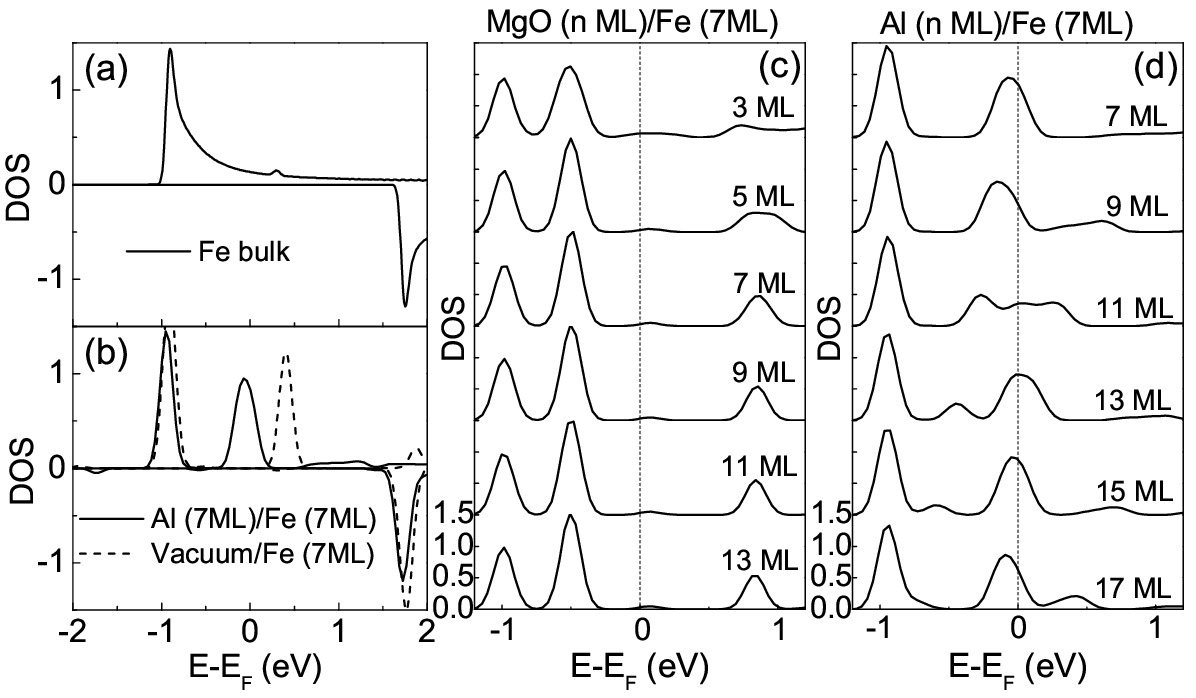}%
\caption{\label{fig1} The DOS of $\Delta_1$ electron at $\Gamma$ point in (a) Fe bulk and (b) middle atomic layer of Fe film in vacuum/Fe(7 ML) structure (dotted line) and Al(7 ML)/Fe(7 ML) multilayer(solid line). The DOS of majority-spin $\Delta_1$ electron at $\Gamma$ point of middle Fe film in (c) MgO(n ML)/Fe(7 ML) and (d) Al(n ML)/Fe(7 ML) multilayers.}
\end{figure}

 Fig. \ref{fig1} shows the density of states (DOS) of $\Delta_1$-symmetry electrons at $\Gamma$ point ($k$$_x$=$k$$_y$=0) in Fe bulk and middle atomic layer of 7ML Fe film. In the energy range of -0.9$\sim$2 eV, the majority-spin DOS of Fe bulk in Fig. \ref{fig1}(a) is smooth, without energy gaps. For vacuum/Fe (7ML) structure, electrons in Fe are fully confined in out-of-plane direction, resulting in discrete energy levels and quantum-well states. As can be seen from the dotted line in Fig. \ref{fig1}(b), the DOS presents several quantum-well peaks and obvious energy gaps. When replacing the vacuum with Al, a similar result can be obtained, although fcc-Al and bcc-Fe do not have a relative bandgap for majority-spin $\Delta_1$-electrons between -0.9eV and 2eV. This is because the reflectivity is energy dependent, even without a bandgap, the interface reflectivity does not suddenly decrease to zero, there is still a finite value \cite{PhysRevLett:156801}. We estimated the interface reflectivity of majority-spin $\Delta_1$ electron at $\Gamma$ point on Fermi level to be $\sim$0.433 through transport calculations of the Al/Fe system. It should be noted that the reflectivity should be squared to a smaller value ($\sim$0.187) for metal multilayers. As shown in Fig. \ref{fig1}(b), with such a small reflectivity, the Al(7ML)/Fe(7ML) multilayers can still generate quantum-well peaks and obvious energy gaps.

 Compared with the full confinement system (vacuum/Fe or MgO/Fe), the partial confinement (Al/Fe multilayer) has several major characteristics. Firstly, except the peak at the band edge, the width of the quantum-well peaks widens and the height decreases, by comparing the solid and dotted lines in Fig. \ref{fig1}(b). Secondly, the $\Delta_1$ wave function will not decay exponentially inside the Al layer as in a full confinement system. This allows Al to not only play a role in generating quantum-well resonance, but also serve as an electrode material. Lastly, the quantum-well levels of Fe layer can be adjusted by both Fe and Al thickness. According to the quantization condition 2$kd$+2$\varphi$=2$\pi$$n$ \cite{PhysrevB:233408,PRApplied:054042}, the quantum-well levels of Fe can be controlled by $d$ (thickness of Fe) and $\varphi$ (interface phase shift), where $n$ is an integer, and $k$ is the perpendicular wave vector for energy $E$ in Fe band. For a full confinement system, $\varphi$ near the Fermi energy is related to the type of barrier materials, but usually does not change with the barrier thickness. As shown in Fig. \ref{fig1}(c), with the change of MgO thickness, the quantum-well peak near the Fermi energy remains almost unchanged, which is consistent with literatures \cite{PhysrevB:134404,JKPS:972}. However, for the partial confinement system, $\varphi$ can be altered by the thickness of Al layer, resulting in the change of quantum-well levels in Fe. As shown in Fig. \ref{fig1}(d), with the increase of Al thickness, the quantum-well peaks move regularly towards a lower energy until it converges toward the bottom of the band (-1eV). This is because the $\Delta_1$ electron in Al film can also be partially confined by two Al/Fe interfaces, and change periodically with Al thickness. The $\varphi$ for quantum well of Fe is largely determined by the electronic structure of Al \cite{PhysRevLett:156801}, and therefore undergoes periodic changes. Ultimately, the quantum-well levels of the Fe layer can be periodically regulated by Al thickness. The predicted additional degree of freedom in partial confinement case may be relatively easily confirmed by inverse photoemission and photoemission experiments.

 In the above discussion, we only focused on the quantum-well resonances of $\Delta_1$ electrons in the $\Gamma$ point. If we consider all symmetric electrons and the entire two-dimensional Brillouin zone (2DBZ), the quantum-well signal caused by partial confinement will be greatly weakened or even disappear due to averaging. In order to electrically demonstrate this quantum-well resonances, we use MgO as the barrier layer of MTJ, which has the lowest decay rate for $\Delta_1$-symmetry electrons near the $\Gamma$ point \cite{PhysRevLett:1088}.

\begin{figure}[thb!]
\centering
\includegraphics[width=8.6cm]{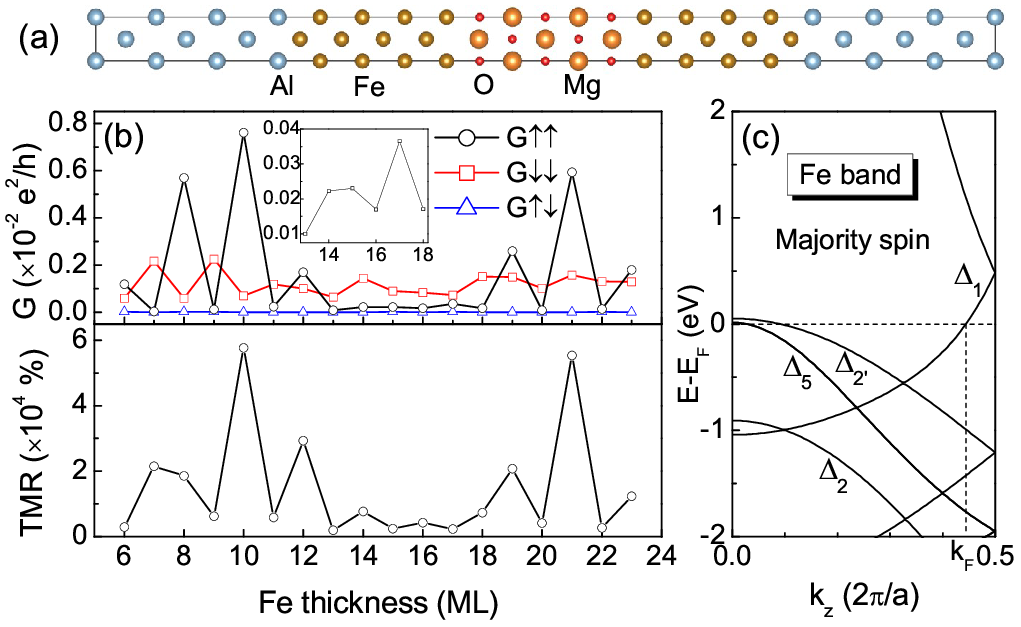}%
\caption{\label{fig2} (Color online) (a) Relaxed structure, (b) conductance and TMR ratio of Al/Fe/MgO/Fe/Al MTJ. (c) Band structure of majority-spin bcc-Fe. The conductance is calculated at the Fermi energy. The small fluctuation areas (13$\sim$18 ML)of $G_{\uparrow\uparrow}$ is enlarged in (b). k$_F$ is the perpendicular Fermi wave vector in bulk Fe.}
\end{figure}

 Fig. \ref{fig2}(a) shows one of the relaxed structures of Al/Fe/MgO/Fe/Al MTJ. MgO barrier is 5 ML ($\sim$1nm). Two Fe layers are fixed to the same thickness. In Fig. \ref{fig2}(b), the spin-dependent conductance and the TMR ratio are shown as a function of Fe thickness. The majority-spin conductance in parallel states ($G_{\uparrow\uparrow}$) oscillates regularly with the change of Fe thickness. From 6 ML to 23 ML Fe, there are 8 cycles for $G_{\uparrow\uparrow}$. So, the oscillation period of $G_{\uparrow\uparrow}$ is estimated to be 2.13 ML Fe. The $ab$ $initio$ calculations by C. Heiliger $et$ $al.$ gives similar oscillation behavior caused by quantum-size effects in Cu/Fe/MgO/Fe/Cu MTJ with changing Fe thickness \cite{PhysRevLett:066804}. Considering the lowest decay rate of $\Delta_1$-symmetry electrons in MgO, the $G_{\uparrow\uparrow}$ value and its oscillation period are mainly determined by the $\Delta_1$ electrons. Because the oscillation behavior comes from the quantum-well resonances, the oscillation period is related to the perpendicular Fermi wave vector $k$$_F$ given by the band of bcc-Fe unit cell. From Fig. \ref{fig2}(c), $k$$_F$ of $\Delta_1$ electrons is about 0.445$\cdot$2$\pi$/$a$ (where $a$ is 2 ML Fe), resulting in a period $T$=$\pi$/$k$$_F$=2.25 ML Fe, which is consistent with the period obtained for $G_{\uparrow\uparrow}$ and in agreement with the literature \cite{PhysrevB:134404}. It should be noted that the oscillation period (2.13ML or 2.25ML) is too short and not an integer, resulting in insufficient density of data points when selecting points in the integer ML. So, the $G_{\uparrow\uparrow}$ value in some thickness ranges is very small, making the number of periods unclear. When counting the number of cycles, the oscillations of small fluctuation areas (13$\sim$18 ML) should be taken into account in order to match the actual situation.

 The TMR ratio in Fig. \ref{fig2}(b) also exhibits corresponding oscillation, but due to the interference of minority-spin conductance in parallel states ($G_{\downarrow\downarrow}$) and the conductance in antiparallel states ($G_{\uparrow\downarrow}$, $G_{\downarrow\uparrow}$), its oscillation period is not as clear as $G_{\uparrow\uparrow}$.

 As a comparison, we also calculated the $G_{\uparrow\uparrow}$ in Co/Fe/MgO/Fe/Co MTJ with an ultralow Co/Fe interface reflectivity $\sim$0.001 of majority-spin $\Delta_1$ electrons near $\Gamma$ point. With changing Fe thickness (from 7$\sim$10 ML), the $G_{\uparrow\uparrow}$ fluctuates slightly around the $G_{\uparrow\uparrow}$ value of Fe/MgO/Fe MTJ ($5.09\times10$$^{-4}$ e$^2$/h), ranging from $4.90\times10$$^{-4}$ e$^2$/h to $6.29\times10$$^{-4}$ e$^2$/h. This indicates that quantum-well resonance of $\Delta_1$ electrons disappears when its interface reflectivity is very low. The slight fluctuation of $G_{\uparrow\uparrow}$ should originate from the quantum confinement of $\Delta_{2'}$- and $\Delta_5$-symmetry electrons in Fe layers.

\begin{figure}[thb!]
\centering
\includegraphics[width=8.6cm]{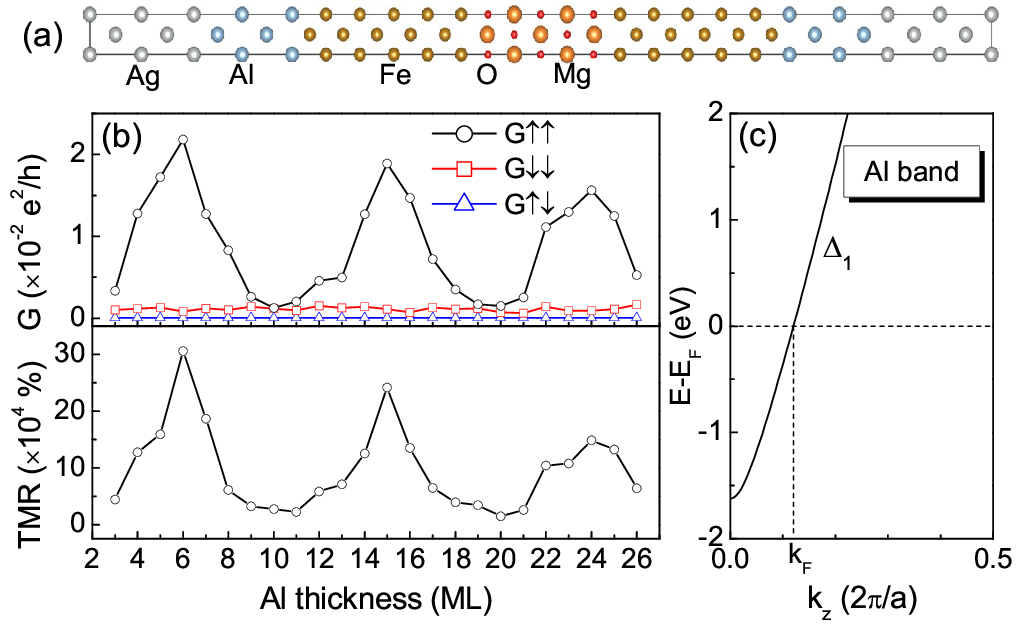}%
\caption{\label{fig3} (Color online) (a) Relaxed structure, (b) conductance and TMR ratio of Ag/Al/Fe/MgO/Fe/Al/Ag MTJ. (c) Band structure of fcc Al. k$_F$ is the perpendicular Fermi wave vector in bulk Al, and changes little near $\Gamma$ point. The band structure is calculated by two Al atoms.}
\end{figure}

 When utilizing the specificity of partial confinement - its ability to adjust quantum-well levels by Al thickness, we constructed the Ag/Al/Fe/MgO/Fe/Al/Ag MTJ, as shown in Fig. \ref{fig3}(a). The thicknesses of the two Fe films are both fixed at 10 ML. The Ag at both ends of the tunnel junction is used as electrodes due to an appreciable reflectivity of $\sim$0.361 at Ag/Al interface, and it should be noted that the two Ag electrodes can be replaced with other metals. The two Al layers have the same thickness. When the thickness of two Al films is changed from 3 ML to 26 ML, the $G_{\uparrow\uparrow}$ oscillates with a period $\sim$9 ML Al, as shown in Fig. \ref{fig3}(b). The exact oscillation period caused by quantum-well resonances can be obtained through the band structure. From Fig. \ref{fig3}(c), the perpendicular Fermi wave vector $k$$_F$ of fcc Al is about 0.12$\cdot$2$\pi$/$a$ (where $a$ is 2 ML Al), resulting in a period $T$=$\pi$/$k$$_F$=8.33 ML Al, which is close to the period obtained from $G_{\uparrow\uparrow}$.

 Compared with Fig. \ref{fig2}(b), the oscillation of $G_{\uparrow\uparrow}$ in Fig. \ref{fig3}(b) is clearer and more regular, with multiple thicknesses included in each oscillation cycle. This reflects the fine tuning of quantum-well levels by Al thickness. In addition, the maximum of $G_{\uparrow\uparrow}$ in Fig. \ref{fig3}(b) is about 2.87 times that of $G_{\uparrow\uparrow}$ in Fig. \ref{fig2}(b), mainly due to the presence of Al thin film increasing the peak value of Fe layer quantum well. Comparing the DOS of Al (59ML)/Fe (7ML) and Al (7ML)/Fe (7ML) multilayers, there is a difference of approximately 1.7 times in the peak value. After squared, it is 2.89, which is exactly close to 2.87. The increase in peak value is originated from the enhanced reflectivity at the Al/Fe interface when Al becomes a thin layer. Therefore, the Al thin films not only enables fine tuning of Fe quantum wells, but also enhances the TMR effect, which breaks through our usual cognition that the metal layers other than the FM layer have a weak impact on the TMR effect.

\begin{figure}[thb!]
\centering
\includegraphics[width=8.6cm]{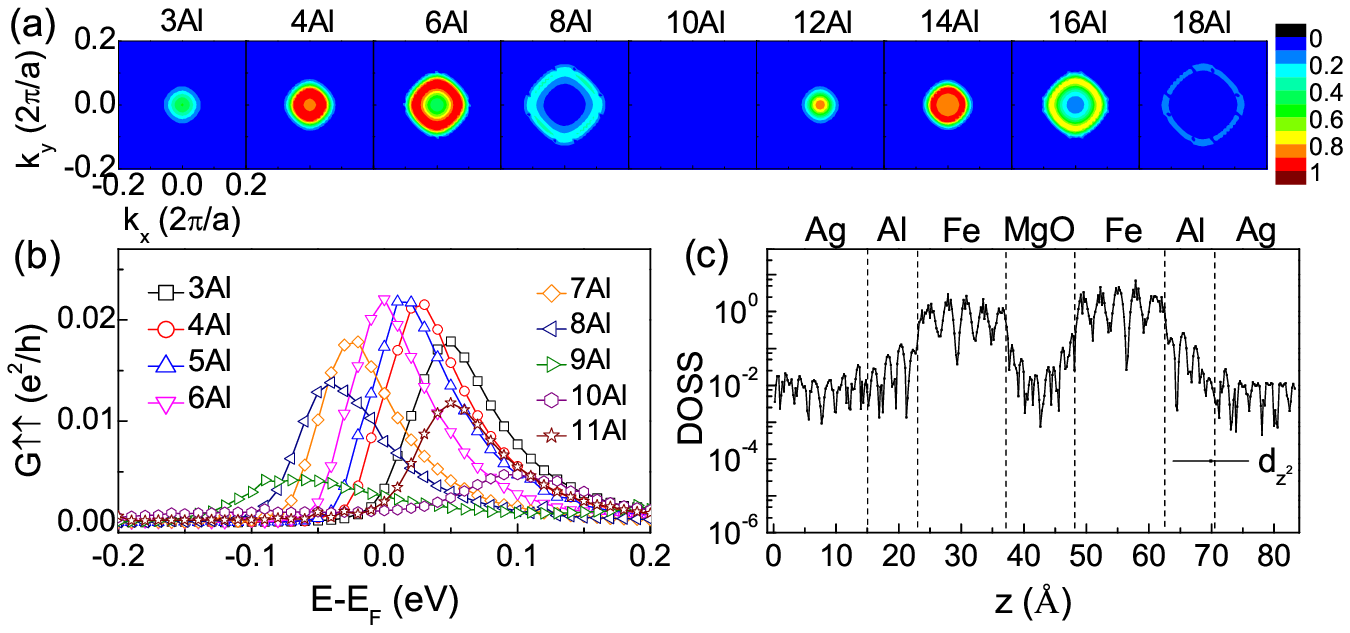}%
\caption{\label{fig4} (Color online) (a) The $k_{\|}$-resolved transmission coefficients for $G_{\uparrow\uparrow}$ at the Fermi energy near the $\Gamma$ point across the Ag/Al/Fe/MgO/Fe/Al/Ag MTJ with changing Al thickness from 3 ML (3Al) to 18 ML (18Al). (b) $G_{\uparrow\uparrow}$ with different Al thickness as a function of energy. (c) DOSS of d$_{z^2}$ orbital at Fermi energy at $\Gamma$ point of MTJ with 4 ML Al in (a) along the transport direction.}
\end{figure}

 In order to further understand the influence of Al thickness on $G_{\uparrow\uparrow}$, we calculated the $k_{\|}$-resolved transmission coefficients and energy dependent $G_{\uparrow\uparrow}$ of the Ag/Al/Fe/MgO/Fe/Al/Ag MTJ with changing Al thickness. The transmission coefficients distribution for $G_{\uparrow\uparrow}$ in Fig. \ref{fig4}(a) reveals a circle and ring feature with periodic changes. As the thickness of Al increases, the circle becomes ring. Then the ring gradually becomes larger, and disappears when it is further away from $\Gamma$ point. Continuing increasing the Al thickness, the circle and ring reappears and then the ring becomes larger. This variation is somewhat consistent with the photoemission data and calculated results in systems with quantum-well resonances \cite{PhysRevLett:156801,PhysrevB:085327,PhysrevB:L060101}. That is, changing the relative position of quantum-well levels and energy point can make in-plane wave vector of the quantum-well peak regularly increases or decreases (change in circle or ring size). Since the energy is fixed at Fermi energy, the circle and ring variation with Al thickness in Fig. \ref{fig4}(a) should come from the shift of quantum-well level. The energy dependent $G_{\uparrow\uparrow}$ further confirmed this point. As shown in Fig. \ref{fig4}(b), with the increase of Al thickness, the $G_{\uparrow\uparrow}$ peaks move regularly towards the lower energy positions. When a peak moves away from the Fermi level, a new peak appears above the Fermi level and continues moving towards lower energy positions. This periodic change is in agreement with the movement of quantum-well peaks of Fe film in Fig. \ref{fig1}(d). Therefore, the variation of $G_{\uparrow\uparrow}$ originates from the modulation of Al thickness to quantum-well level of Fe film. This characteristic of partial confinement case can achieve the coincidence of quantum-well level and Fermi level, which is beneficial for MTJs that can only apply small bias.

\begin{table*}
\renewcommand\arraystretch{1.2}
\caption{\label{table1} Spin-dependent Transmission at Fermi level of four-types MTJs. $G_{\uparrow\uparrow}$, $G_{\downarrow\downarrow}$, $G_{\uparrow\downarrow}$ and $G_{\downarrow\uparrow}$ are in units of e$^2$/h. The parallel ($G_P$=$G_{\uparrow\uparrow}$+$G_{\downarrow\downarrow}$) and antiparallel conductance ($G_{AP}$=$G_{\uparrow\downarrow}$+$G_{\downarrow\uparrow}$) define the optimistic and normalized TMR ratio to be ($G_P$-$G_{AP}$)/$G_{AP}$ and ($G_P$-$G_{AP}$)/($G_P$+$G_{AP}$), respectively. System \uppercase\expandafter{\romannumeral1}, \uppercase\expandafter{\romannumeral2}, \uppercase\expandafter{\romannumeral3} are Al/Fe(10)/MgO/Fe, Al/Fe(10)/MgO/Fe(10)/Al and Ag/Al(6)/Fe(10)/MgO/Fe(10)/Al(6)/Ag MTJ (unit: ML), respectively.}
\begin{ruledtabular}
\begin{tabular}{cccccccc}
MTJ & $G_{\uparrow\uparrow}$ & $G_{\downarrow\downarrow}$ & $G_{\uparrow\downarrow}$ & $G_{\downarrow\uparrow}$ & Optimistic TMR (\%) & Normalized TMR (\%) \\
\cline{1-7}
Fe/MgO/Fe & $5.09\times10$$^{-4}$ & $2.69\times10$$^{-4}$ & $1.27\times10$$^{-5}$ & $1.32\times10$$^{-5}$ & $2.90\times10$$^3$ & 93.53 \\
System \uppercase\expandafter{\romannumeral1} & $1.84\times10$$^{-3}$ & $2.19\times10$$^{-4}$ & $8.57\times10$$^{-6}$ & $1.36\times10$$^{-5}$ & $9.19\times10$$^3$ & 97.87 \\
System \uppercase\expandafter{\romannumeral2}& $7.62\times10$$^{-3}$ & $4.65\times10$$^{-4}$ & $7.21\times10$$^{-6}$ & $7.13\times10$$^{-6}$ & $5.63\times10$$^4$ & 99.65 \\
System \uppercase\expandafter{\romannumeral3} & $2.18\times10$$^{-2}$ & $8.21\times10$$^{-4}$ & $3.76\times10$$^{-6}$ & $3.65\times10$$^{-6}$ & $3.05\times10$$^5$ & 99.93 \\
\end{tabular}
\end{ruledtabular}
\end{table*}

 In Fig. \ref{fig4}(a), the transmission coefficient of the bright red area is greater than 0.9. Since only one $\Delta_1$ band crossing the Fermi level plays a major role in transport, the transmission coefficient divided by the number of bands results in a transmissivity of greater than 0.9. The highest transmissivity in these bright red areas is 0.999. We took the $\Gamma$ point (transmissivity $\sim$0.875) of MTJ with 4 ML Al in Fig. \ref{fig4}(a) and calculated its density of scattering states (DOSS). Fig. \ref{fig4}(c) shows the DOSS of d$_{z^2}$ orbital. The d$_{z^2}$ ($\Delta_1$-symmetry) electrons hardly decays throughout the transport process, proving that the existence of quantum-well resonance tunneling allows electrons to pass through the barrier layer almost without loss. It should be pointed out that the existence of quantum-well resonances does not necessarily mean a transmissivity of $\sim$1. For example, in Al/Fe/MgO/Fe/Al MTJ, the highest transmissivity is $\sim$0.4 in the 2DBZ. For another example, in Ag/Al/Fe/MgO/Fe/Al/Ag MTJ with thicker MgO (7 ML MgO), the highest transmissivity is $\sim$0.145 in the 2DBZ. When the resonance tunneling signal is not strong enough or the barrier effect is too large, non-attenuating tunneling will not occur.

 Table \ref{table1} lists the transport properties of several MTJs. The optimistic TMR ratio in MTJs with quantum-well resonances is much higher than the Fe/MgO/Fe MTJ, mainly due to the increase of $G_{\uparrow\uparrow}$ caused by quantum-well resonances. It is worth noting that even in the Al/Fe/MgO/Fe MTJ with single quantum well, there is an obvious improvement in the transport properties compared to Fe/MgO/Fe MTJ, which is beneficial for achieving enhancement of TMR ratio in practice.

 We also tested the effects of strain and small interface reflectivity. When we apply 1\% or 2\% strain, similar conductance oscillations and high TMR effect can be obtained. Strain will slightly alter the band structure and affect the oscillation period, resulting in a change in the thickness of Al and Fe where the highest TMR ratio is located. In addition, when we replace fcc Al with fcc Ag, fcc Au, fcc Re or fcc W, the predicted quantum-well resonances can be reproduced. Through comparison these structures, we found that a small interface reflectivity will reduce the amplitude and the maximum of TMR oscillation. Among the calculated MTJs, the improvement effect of the Al system is better.

 In the end, we have to say that it is difficult to ensure that the quantum-well levels on both sides of MgO coincide in pracitce, which may greatly weakens TMR effect and goes against expectations. Fortunately, as mentioned above, even Al/Fe/MgO/Fe MTJ with single quantum well can significantly enhance the TMR effect. This gives us a lot of space to design MTJ with quantum-well resonances. The work will promote the development of quantum-well resonance in spintronics devices, especially in improving the performance of MRAM and magnetic sensors.

\section{ACKNOWLEDGEMENT}
 This work was supported by the National Key Research and Development Program of China [MOST, Grant No. 2022YFA1402800], the National Natural Science Foundation of China [NSFC, Grant Nos.12134017, and 12204517], and partially supported by the Strategic Priority Research Program (B) [Grant No. XDB33000000]. The atomic structure visualisation was produced with VESTA software \cite{vesta}.

\end{document}